\documentclass[structabstract]{aa}  
\usepackage{graphicx}
\usepackage{txfonts}
\usepackage{natbib}
\begin{document}

   \title{Upsilon Andromedae b in polarized light:
          New constraints on the planet size, density and albedo}
%   \subtitle{}
   \titlerunning{Upsilon Andromedae b in polarized light}

   \author{S. V. Berdyugina   \inst{1}
          \and
          A. V. Berdyugin     \inst{2}
          \and
          V. Piirola          \inst{2}
    }
    
   \institute{Kiepenheuer Institut f\"ur Sonnenphysik, 
              Sch\"oneckstrasse 6, D-79104 Freiburg, Germany\\
              \email{sveta@kis.uni-freiburg.de}
          \and
              Finnish Centre for Astronomy with ESO (FINCA), University of Turku,
              V\"ais\"al\"antie 20, FIN-21500, Piikki\"o, Finland\\ 
              \email{andber@utu.fi, piirola@utu.fi}
    } 
    
    \date{Received ...; accepted ...}
    
% \abstract{}{}{}{}{} 
% 5 {} token are mandatory
\abstract
  % context heading (optional) leave it empty if necessary 
   {  Polarimetry is a novel tool to detect and characterize exoplanets and their atmospheres. It provides unique constraints on the planet orbit parameters, such as its inclination, orientation in space and shape, as well as reflective properties and composition of the atmosphere. These combined with spectroscopic and photometric measurements can fully characterize the planet with respect to its size, density, temperature, composition, etc., even for non-transiting systems. }  
   { Polarized scattered light from the non-transiting hot Jupiter $\upsilon$~And~b is measured to further constrain its orbit, mass, density, and geometrical albedo. }
  {We obtained polarimetric measurements in the $UBV$ bands over the orbital period and  deduce an average peak-to-peak amplitude of $(49 \pm 5)\times10^{-6}$ in both Stokes $q$ and $u$.}
  {From our data we evaluate the orbit inclination $i=111\degr\pm11\degr$, longitude of the ascending node $\Omega=236\degr\pm12\degr$ (or equivalently 56\degr), the effective size of the scattering atmosphere in the optical blue of $1.36\pm0.20$\,$R_{\rm J}$. These combined with spectroscopic measurements result in the planet mass $0.74\pm0.07$\,$M_{\rm J}$, mean density $0.36\pm0.08$\,g\,cm$^{-3}$, and surface gravity $\sim10^3$\,cm\,s$^{-2}$, which favor a close similarity of $\upsilon$~And~b to other inflated hot Jupiters. We also significantly improved the periastron epoch $T_{\rm p}={\rm JD}2,450,032.451$, interior conjunction epoch $T_{\rm t}={\rm JD}2,450,034.668$, and periastron longitude $\omega=279\degr\pm14\degr$. The latter indicates that the apsidal resonance known for planets c and d includes also planet b.
  Obtained limits on the wavelength dependent geometrical albedo (average 0.35) indicate its similarity to Neptune with peak reflectivity in the blue. Combinig all available measurements at various passbands, we construct a unified wavelength dependent albedo of an average hot Jupiter. It appears to be largely shaped by Rayleigh scattering in the blue and atomic and molecular absorption in the optical and near infrared.
  }
  {Our findings demonstrate the power of polarimetry for studying non-transiting exoplanets.}

   \keywords{planets and satellites: detection --- 
             planets and satellites: atmospheres ---
             planets and satellites: fundamental parameters ---
             polarization --- 
             techniques: polarimetric --
             planets and satellites: individual: Upsilon Andromedae b
            }
            
\maketitle
            
%________________________________________________________________

\section{Introduction}

Upsilon Andromedae b ($\upsilon$~And b) is a short period (4.6 days), non-transiting hot Jupiter orbiting an F8 dwarf together with other three planets \citep{but97,but99,cur11}. Planets in such systems are often found close to resonance orbits which is highly interesting for our understanding of planet formation and evolution. For instance, in the $\upsilon$~And system planets c and d are in apsidal alignment associated with a secular resonance \citep{chi01}, while planet e is close to an external 3:1 resonance with planet c \citep{cur11}. Astrometric measurements of planets c and d with {\it Hubble Space Telescope} combined with radial velocity variations revealed their very low orbit inclinations and high masses \citep{mca10}. Their mutual orbit inclination of 30\degr\ appears to be at the limit for this system to be stable \citep[e.g.,][]{mic06,bar11}. 

Photons interacted with or emitted by an exoplanet are a key source of information on its physical properties. So far, our knowledge relies mainly on transition spectroscopy and secondary eclipse techniques and, therefore, is largely limited to transiting planets. Exceptions are $\upsilon$~And~b and a few other hot Jupiters, for which a variable infrared flux was measured directly \citep[e.g.,][]{har06,cross10}, and HD189733b, directly studied with the help of optical polarimetry \citep[][hereafter B08 and B11]{berd08,berd11}. The latter studies further constrained orbital parameters and revealed a strong wavelength dependence of reflecting properties (geometrical albedo) of the planet with a maximum in the blue. This was found to be strikingly similar to Neptune which appears bright in the blue due to Rayleigh scattering and absorption in the red \citep{karkoschka1994,sromovsky2005}. Thus, polarimetric studies are essential for non-transiting planets as they provide an inclination of the orbit (hence, true mass), radius of the planet (hence, density), and direct access to planetary atmospheres \citep{seager00,stam04,flu10,berd11spw6}. 

This paper reports measurements of the reflected light from a non-transiting extrasolar planet using polarimetry in the $UBV$ bands. Form these data we evaluate the inclination and orientation of the $\upsilon$~And~b orbit and the planet radius, which result in reliable estimates of its mass and mean density. In addition, we find indications of an atmosphere which efficiently scatters the stellar light in the blue ($UB$ bands). Previous polarimetric observations of $\upsilon$~And~b by \citet{hou06} were made in a broad red band (covering $RI$ bands), and no detection or limits were reported, which implied a very low albedo in the red. 
%An upper limit on a planet radius was deduced from a probable spectral signature of the reflected light \citep{cc02}. 
We estimate an average optical albedo of $\upsilon$~And~b of 0.35, similar to that of HD189733b (B11) and Kepler-7b \citep{kip11,dem11}. These three planets are inflated and highly reflective hot Jupiters, most probably due to Rayleigh scattering on condensates in high altitude haze or clouds \citep{berd11spw6, dem11}.

%***********************************************************************
\section{Observations}\label{sec:obs}  
%***********************************************************************

The observations were carried out in 2009 October 27 to November 3 and in 2010 September 15--22 with the  TurPol broad-band ($UBVRI$) polarimeter \citep{piirola73,piirola88} mounted in the Cassegrain focus at the 2.5m Nordic Optical Telescope (NOT). The observing procedure was the same as in B11. In particular, the Stokes $qu$-system was defined in accordance with the common agreement that positive and negative $q$ are in the north-south and east-west directions, respectively, while positive and negative $u$ are at an angle of 45\degr\ counter-clockwise from the positive and negative $q$.
 
To measure linear polarization (normalized Stokes $qu$), the polarimeter is equipped with the following elements along the optical path: (1) a half-wave plate retarder rotating at 22.5\degr\ intervals to modulate between polarization states, (2) calcite block to split the light onto the ordinary '$o$' and extraordinary '$e$' parallel light beams, (3) two diaphragms to limit the light integration area around each beam, (4) chopper to quickly (25 Hz) switch between the ordinary and extraordinary beams, (5) dichroic mirrors to separate the passbands, (6) field lenses to direct each of the two beams on the same detector, and (7) five photomultipliers as detectors for the five $UBVRI$ bands. A drawing of the principle scheme for one channel can be found in \citet{piirola73}. 

There are a few important advantages in this scheme which enable achieving very precise polarimetry with TurPol.

First, since the calcite block is located {\it above} the diaphragms, an additional $e$-component from sky background, (sky)$_e$, is superimposed to the (star+sky)$_o$ component, while an additional (sky)$_o$ is sperimposed to (star+sky)$_e$. Thus, {\em both} components of sky polarization pass {\em both} diaphragms.
Since they are of the opposite polarization states, they cancel out exactly, independently on the amount and origin of background polarization. Therefore, polarization due to, e.g., scattering in the Earth atmosphere, moonlight, or resolved extended sources (nebulae) is exactly eliminated at the time of target observations. This is a great advantage of TurPol in comparison with the polarimeters PlanetPol of \citet{hou06} and its copy POLISH of \cite{wik09} where a Wollaston prism is used instead of a simple calcite block. The prism splits the light with a large divergence angle between the ordinary and extraordinary beams, and therefore the sky beams do not overlap, and their polarization does not cancel out. Therefore, such polarimeters should measure very carefully background fluxes in each beam separately to deduce the sky polarization and subsequently subtract it from target measurements. This is a significant source of errors in such schemes. For instance, \citet{luc09} reported that dust blown from Sahara to Canary Islands can produce sky polarization of order 10$^{-5}$ at large zenith distances. The compensation scheme of TurPol guaranties that this spurious polarization is completely eliminated at the time of measurements.

Second, the rapidly rotating chopper (25\,Hz) alternates integration of the ordinary and extraordinary beams on the same detector. This strongly diminishes systematic effects due to instrumental sensitivity (flat-field effects), seeing and variable sky transparency. Again, this is an advantage to other polarimeters where integration of orthogonal polarization states is carried out with different detectors and at much longer modulation times. 

Third, thanks to the dichroic mirrors and five detectors, several passbands can be measured simultaneously. This significantly increases the efficiency of observations and provides information on spectral distribution of polarization, which is valuable for identification of its source.

Finally, each pair of the observed Stokes $q$ and $u$ was calculated from eight exposures at different orientations of the retarder, which allowed us to avoid systematic errors due to imperfections of the retarder. 

To calibrate the data with high accuracy, we dedicated a significant amount of observing time for determining systematic effects. For this purpose, we observed 10 nearby, bright, nonvariable, and nonpeculiar stars ($V$=5$^{\rm m}$--6$^{\rm m}$, $d<50$\,pc) of spectral classes F--G. This range allowed us to maximize the photon flux in the blue and observe without neutral density filters. The stars were observed on the same nights as $\upsilon$~And at various parallactic angles to determine instrumental polarization which includes contributions from the telescope ($P_{\rm tel}$) and the polarimeter itself ($P_{\rm in}$). The former is rotating during observations due to the alt-azimuthal mount and the latter is constant. Total integration time of 3--4 hours for each star allowed for accuracy of (1--2)$\times$10$^{-5}$. All stars, including scientific targets, were observed within $\pm$30\degr\ zenith distance to minimize variations of instrumental and sky polarization and to maximize the photon flux.
In addition, for calibration of the polarization angle zero point, we observed highly polarized standard stars HD25443, HD161056, and HD204827, with the accuracy better than 1\degr.

The polarization calibration data are presented in Table~\ref{tab:stand}. $P_{\rm tel}$ corresponds to the telescope position angle zero. Rotational transformation of these values are applied as a function of the angle of the telescope relative to the equatorial system, due to the alt-azimuthal mounting.

\begin{table*}
\caption{Polarization calibration data. Stokes parameters $[q, u]\pm\sigma$ are given in units of 10$^{-5}$. 
} 
\label{tab:stand}
\centering
\begin{tabular}{llrrrr}
\hline\hline
Target & Sp. & $U$ & $B$ & $V$ & $UBV$ \\
\hline
\multicolumn{5}{c}{October 2009}\\
% October 2009
HD24740  & F2 IV &  [2.1, --2.7] $\pm$ 3.1  &   [--1.3, --3.5] $\pm$ 2.3   &  [--5.7,  2.1] $\pm$ 4.1  & [--1.0, --2.3] $\pm$ 2.2\\ % V=5.6 pi=22
HD29645  & G0 V  & [--5.0,--2.5] $\pm$ 2.1  &   [--0.7, --1.0] $\pm$ 1.4   &  [0.6,  --2.8] $\pm$ 3.2  & [--1.7, --1.6] $\pm$ 1.6 \\ % V=6.0 pi=32
HD33167  & F5 V  & [--2.2,  3.2] $\pm$ 2.9  &   [--4.2,   1.6] $\pm$ 1.9   &  [--1.8, --7.5] $\pm$ 4.4 &  [--3.4, 1.0] $\pm$ 2.2\\ % V=5.7 pi=22
HD35296  & F8 V  &   [3.9,  1.0] $\pm$ 2.0  &       [3.4, 1.8] $\pm$ 1.4   &    [2.8,  5.1] $\pm$  2.7 &  [3.4, 2.1] $\pm$ 1.4\\ % V=5.0 pi=68
%----------- Instrumental -----------
$P_{\rm in}$    &&   [2.5,--1.3] $\pm$ 1.2 &  [--0.5,--2.8] $\pm$ 0.8  &  [6.0,--3.8] $\pm$ 1.7 & [1.2, --2.5] $\pm$ 0.9\\
$P_{\rm tel}$   && [--1.4,--2.3] $\pm$ 1.2 &    [0.2,--1.1] $\pm$ 0.8  &  [--0.8,1.6] $\pm$ 1.7 & [--0.4, --1.1] $\pm$ 0.9\\
\hline
\multicolumn{5}{c}{September 2010}\\
% September 2010:
HD16234  & F7 V  & [--4,--1] $\pm$ 3  &     [--2, 2] $\pm$  2   &     [2, 0] $\pm$   3 & [--1.5, 0.8] $\pm$ 1.8 \\ % V=5.7 pi=28
HD29645  & G0 V  &  [9, --2] $\pm$ 5  &   [--2,   1] $\pm$  4   &  [--4,  4] $\pm$   3 & [--1.0, 2.0] $\pm$ 2.8 \\ % V=6.0 pi=32
HD159332 & F6 V  &  [--2, 0] $\pm$ 3  &     [0, --5] $\pm$  3   &  [--1,  0] $\pm$   4 & [--1.0, --2.0] $\pm$ 2.3 \\ % V=5.7 pi=27
HD182807 & F7 V  &  [--3, 6] $\pm$ 9  &     [4,  14] $\pm$ 10   & [--10,--4] $\pm$  12 & [--2.3, 6.4] $\pm$ 7.3 \\ % V=6.2 pi=36
HD185395 & F4 V  &    [8, 4] $\pm$ 5  &       [6, 0] $\pm$  5   &     [5, 0] $\pm$   6 & [6.5, 1.5] $\pm$ 3.8 \\ % V=4.5 pi=54
HD225239 & G2 V  &    [1, 1] $\pm$ 4  &       [1, 0] $\pm$  3   &   [3, --3] $\pm$   3 & [1.8, --1.0] $\pm$ 2.4 \\ % V=6.1 pi=27
%------- Instrumental -------------
$P_{\rm in}$    && [1.5,--5.1] $\pm$ 1.6 & [--2.7,--2.9] $\pm$ 1.4  &    [0.0,--0.1] $\pm$ 1.6 & [--0.6, --2.7] $\pm$ 1.1 \\
$P_{\rm tel}$   && [--5.2,0.9] $\pm$ 1.6 &   [--4.1,0.1] $\pm$ 1.4  &    [--0.7,0.1] $\pm$ 1.6 & [--3.4, 0.3] $\pm$ 1.1 \\
\hline
\end{tabular}
\end{table*}

After the re-aluminization of the primary mirror of the NOT in spring 2009 $P_{\rm tel}$ reduced by about a factor of 10 as compared to 2008 (see B11) and became comparable with measurement errors. In 2010 it slightly increased but still remained significantly lower than in 2008. This is also confirmed by measurements in June 2010 during another observing run with TurPol. {  $P_{\rm in}$ may include small effects from residual interstellar polarization of the standard stars, although these effects tend to average out for a sample of stars in different directions on the sky. In any case, uncertainties of $P_{\rm in}$ will only cause a constant shift of the $q$- and $u$-curves, and have no effect on the amplitude of the variations we are using for our analysis.}

\begin{figure}
\centering
\resizebox{\hsize}{!}{\includegraphics{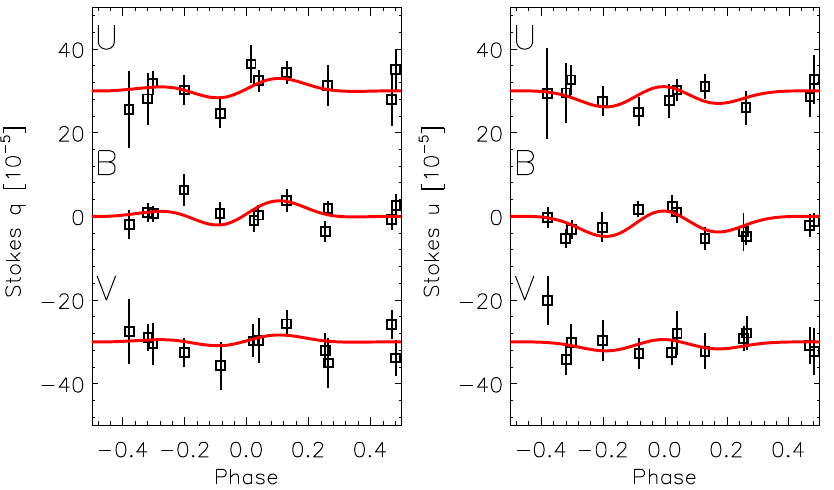}}
\resizebox{\hsize}{!}{\includegraphics{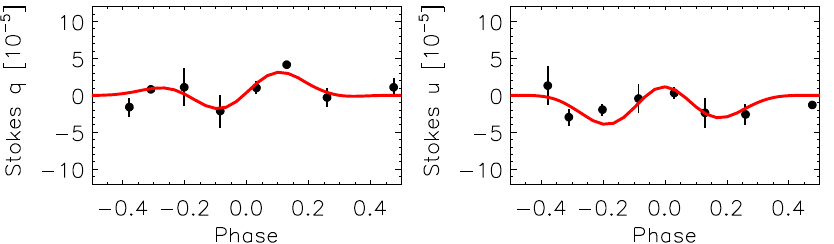}}
%\resizebox{\hsize}{!}{\includegraphics{f1c.pdf}}
\caption{Stokes $q$ and $u$ with $\pm1\sigma$ error bars for $\upsilon$~And~b. 
Upper panels: nightly $UBV$ measurements for each band separately (squares). The $U$ and $V$ data are shifted in vertical by $\pm$3$\times$10$^{-4}$ for clarity. 
Lower panels: the $UBV$ data averaged together. The mean error of the binned data is 7$\times$10$^{-6}$ and the standard deviation is 4.5$\times$10$^{-6}$. 
Curves are best-fit solutions (Sect.~\ref{sec:mod}) obtained by using orbital parameters listed in the last column of Table~\ref{tab:par} for each band individually (upper panel) and all bands simultaneously (lower panel). Phase 0.0 corresponds to the periastron epoch.
}
\label{fig:pol}
\end{figure}

We observed $\upsilon$~And by making 10\,s exposures in the $UBV$ passbands simultaneously, during two hours on each clear nights. The best photon statistics was in $B$ because of the stellar spectral distribution. Since the star is very bright, we used a 10\% neutral density filter which allowed the maximum photon flux in $B$ below the saturation limit. Because of bad weather we lost six nights. Errors of individual measurements obtained during one cycle of the retarder were conservatively assigned to either statistical or photon noise errors, whatever was larger. This may have lead to some overestimation of the errors, as the normalized $\chi^2$ of our fits to the data was about 0.8. By averaging individual data points we obtained nightly measurements (35 in total) with mean individual errors of 4.5$\times$10$^{-5}$ in $U$ and $V$ and 2.5$\times$10$^{-5}$ in $B$. 
The data were corrected for the telescope and instrument polarization. 
The calibrated measurements are shown in Figure~\ref{fig:pol}. The following peak-to-peak amplitudes in the individual $UBV$ bands were evaluated (in the 10$^{-5}$ scale): 
in Stokes $q$, respectively, 
4.6$\pm$1.9,  
5.8$\pm$1.0, 
2.5$\pm$1.8; 
and in Stokes $u$
4.9$\pm$1.7, 
6.1$\pm$1.1, 
2.7$\pm$3.2.
The uncertainties were calculated as standard deviations from the best fits (see Section~\ref{sec:mod}). Thus, a statistically significant signal was measured only in the $B$ band. However, by averaging altogether the $UBV$-band data we can further improve the statistics and determine the Stokes $q$ and $u$ amplitudes, respectively, $(48\pm 4)\times$10$^{-6}$ and $(50\pm 5)\times$10$^{-6}$ (shown in the lower panels of Figure~\ref{fig:pol}). This is the highest accuracy achieved so far in blue wavelengths. Even a conservative error of $\pm 7\times$10$^{-6}$ (which is the mean error of the binned data) suggests the signal to be real. This average error is, however, not used in our subsequent interpretation and provided here only for evaluating the statistics of the data. A Monte Carlo analysis and parameter fits presented in Sect.~\ref{sec:mod} employ the individual nightly measurement errors. 

As an additional test for robustness of the detection, we made a model fit to the standard star data (Table~\ref{tab:stand}) phased with the $\upsilon$~And b period. The maximum peak-to-peak amplitude found in both Stokes $q$ and $u$ was $(2\pm1)\times$10$^{-5}$, which is within the measurement errors and confirms that the evaluated amplitude of $(5\pm0.5)\times$10$^{-5}$ is not due to calibration errors.

%***********************************************************************
\section{Possible sources of polarization}\label{sec:pol}
%***********************************************************************

We interpret the observed polarization as the result of scattering in the atmosphere of planet b which varies as the planet orbits the star. This is the innermost planet in the system, and scattering polarization from the more distant planets in the system is expected to be significantly smaller than 10$^{-6}$. Two peaks per planet period in the phase curves suggest that the signal is due to reflection. Moreover, planet parameters determined from them agree with previous results (see Sect.~\ref{sec:mod}).

There could also be a stellar contribution to the polarization. This can arise due to the Zeeman effect in spectral lines forming in magnetic stellar regions, such as starspots and plages. In order to estimate the amount of polarization due to active regions, we need both the stellar rotation period and the amplitude of variability. Measurements in the chromospheric \ion{Ca}{ii}~H\&K lines indicate a very low activity and variability of the star \cite[e.g.,][]{sim10}. Because of that, its rotational period is poorly known. The \ion{Ca}{ii}~H\&K records reveal weakly significant periods of 7.3 days \citep{sim10}, 11--12 days \citep{hen00,shk08}, and even 18--19 days \citep{hen00}. For the period of 11--12 days to be due to rotation, the stellar radii should be about $2.1-2.3 R_\odot$. This can be excluded based on the estimate of the stellar angular diameter of 1.1 mas from infrared flux measurements \citep{bla90} and the parallax of 74 mas \citep{van07}, leading to the radius of $1.6 R_\odot$, which is also in agreement with the value of $1.64R_\odot$ suggested by \cite{tak07}. A combination of $v\sin i=9.6$ \citep{val05} and $1.6R_\odot$ suggests the stellar rotation period to be shorter than 8.5 days.

\begin{figure}
\centering
\resizebox{\hsize}{!}{\includegraphics{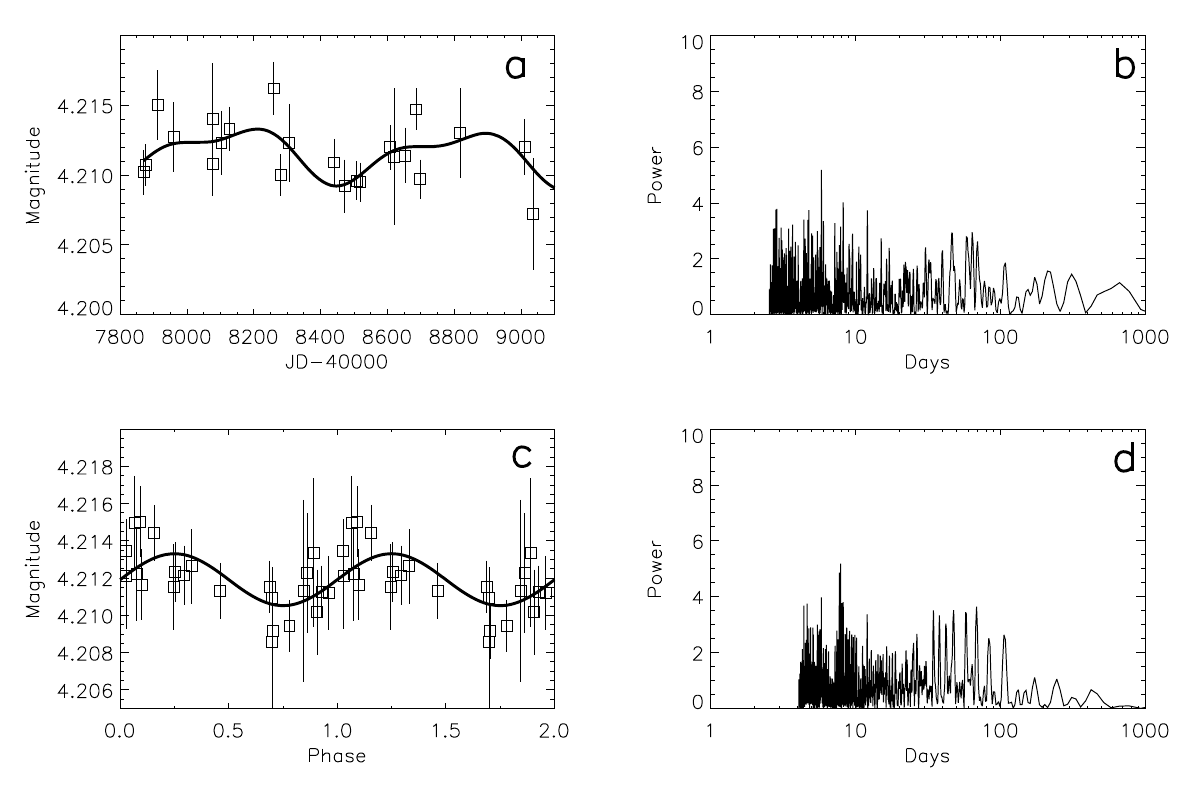}}
\caption{{\it Hipparcos} photometry of $\upsilon$~And. 
{\bf (a)} Original nightly averages (squares) and a double-harmonic fit (curve) 
          revealing a possible long-term modulation of 684 days. 
{\bf (b)} Periodogram of the data from panel a. 
{\bf (c)} Nightly averages from which the long-term modulation was subtracted. 
          The light curve is phased with the 7.88~d period.
{\bf (d)} Periodogram of the data from panel c.
}
\label{fig:phot}
\end{figure}

To obtain additional constraints on the period and to estimate typical projected area of active regions, we analyse here the {\it Hipparcos} photometry \citep{per97}. There are 93 measurements taken during 23 days distributed over about 3 years (Fig.~\ref{fig:phot}). A double-harmonic fit to these data indicates a possible long-term modulation (684 days), while a Lomb-Scargle periodogram suggests a short-term variability of 7.88 days (both periods are of low significance). Variations of the mean stellar brightness on time scales of several years are typical for solar-type spotted stars \citep{ber05}. They can be caused by cyclic growth and decay of active regions as well as by their rearrangement on the stellar surface due to global magnetic field evolution. The short period is close to what is expected for the stellar rotation and to the period of 7.3 days by \cite{sim10}. It is better visible after removing the long-term modulation from the data (see lower panels in Fig.~\ref{fig:phot}). If true, this period implies the stellar rotation axis inclination of 68\degr. Amplitudes of the long-term and rotational modulations are about 0.004 mag and 0.003 mag, respectively. We consider these as upper limits on stellar photometric variability. 

The photometric variability can be caused by either dark spots or bright plages (or both). We estimate that in either case the projected area of an active region causing such a variability is of the order of 0.5\% (slightly depending on the temperature contrast). This is comparable on average with the solar variability. A linear polarization signal from such an active region integrated over the stellar disk and a broad passband would be significantly smaller than $10^{-6}$. Other effects, such as symmetry breaking of the stellar limb polarization due to active regions and stellar oblateness, are even less significant (see, e.g., estimates for HD189733 in B11). Thus, we are safe to assume that the contribution from stellar magnetic regions is negligible.

%***********************************************************************
\section{Orbital parameters}\label{sec:mod}
%***********************************************************************

Modeling observed variations of Stokes $q$ and $u$ enables reconstruction of the orbit spatial orientation and an estimate of the effective size of the scattering atmosphere. Similar to the analysis by B08 and B11, we employ a model based on the Rayleigh--Lambert (RL) approximation for an extended star \citep[][hereafter, FB10]{flu10}, i.e.\ assuming 1) Rayleigh scattering for polarization and 2) the Lambert sphere with the geometrical albedo $A_{\rm g}=$2/3 for intensity. The second assumption has a negligible effect on the phase curve shape for unresolved systems (FB10). 

Since the albedo of the atmosphere is fixed in this model, the wavelength dependent properties of the planet are effectively included into the radius of the RL-atmosphere $R_{\rm RL}$, which represents a geometrical limit for the unity optical thickness in the atmosphere in a given passband. This implies that a possible gas layer above the reflecting surface (e.g., clouds) has no influence on the flux or polarization (i.e., very optically thin). If the radius of the planet is known from other measurements, e.g.\ from transits, the geometrical albedo can be evaluated as in B11. Even though these radii and albedos should only be considered as limits for a spherically symmetric case, they are useful for evaluating fundamental properties of the planet. 

We used the $\chi^2$ minimization procedure described and tested by FB10. Values of parameters previously evaluated from spectroscopy were fixed. These were the orbital period $P$, periastron epoch $T_{\rm p}$, semi-major axis $a$, eccentricity $e$, and the radius of the star $R_*$, which was considered to be a limb-darkened sphere \citep{claret00}. We considered three most accurate sets of parameters obtained by \citet{but06}, \cite{mca10}, and \citet{cur11} (see Table~\ref{tab:par}) and searched for polarimetric solutions for each set separately. Our free parameters were the orbit inclination $i$, longitude of the ascending node $\Omega$, periastron longitude $\omega$, and radius $R_{\rm RL}$. Initially we also varied eccentricity, but the amount of the data was only sufficient to conclude  that a non-zero eccentricity is necessary to achieve better fits. Therefore, we fixed the eccentricity to spectroscopic values. Since the orbit is nearly circular, the maxima are expected near the elongations. The eccentricity results in shifting the maxima closer to the periastron, which constrains $\omega$ and $T_{\rm p}$. The value of the latter was adjusted to place phase 0.0 at the periastron. The epoch of the interior conjunction (transit epoch $T_{\rm t}$ in occulting systems) was assigned to the system brightness minimum. The deduced epochs agree well with those determined by others (Table~\ref{tab:par}). 

\begin{table*}
\caption{Orbital and physical parameters of $\upsilon$~And~b.}
\label{tab:par}
\centering
\begin{tabular}{lcccc}
\hline\hline
Parameter & Value$^{[1]}$ & Value$^{[2]}$ & Value$^{[3]}$ & Value$^{[4]}$ \\
\hline
%-----------------------------------------------------------------------------------
\multicolumn{4}{c}{} \\
P, days             & 4.61711(8) & 4.61711(1)      & 4.61703(3)   & 4.61711$^{[5]}$   \\
$T_{\rm p}^{[6]}$   & 1802.6(7)  & $\dots$         & $\dots$      & 32.451(1)$^{[7]}$ \\ %4.749(?) \\
$T_{\rm t}^{[6]}$   & 1802.97(3) & 34.1(3)         & 5.37(5)      & 34.668(1)$^{[7]}$  \\
$a$, AU             & 0.060(3)   & 0.0594(3)       & 0.0592217(2) & 0.0594$^{[5]}$ \\
$e$                 & 0.02(2)    & 0.012(5)        & 0.0215(7)    & 0.012$^{[5]}$    \\
i\degr\             & $\dots$    & $\dots$         & $\dots$      & 111(11)            \\
$\omega\degr$       & 63(?)      & 44(26)          & 325(4)       & 279(14)            \\
$\Omega\degr$       & $\dots$    & $\dots$         & $\dots$      & 236(12)$^{[8]}$    \\
$M/M_{\rm J}\sin i$ & 0.69(6)    & 0.69(2)         & 0.688(4)     & $\dots$           \\
$M/M_{\rm J}$       & $\dots$    & $\dots$         & $\dots$      & 0.74(7)          \\
$R_{\rm RL}$/$R_{\rm J}$ & $\dots$ & $\dots$  & $\dots$      & 1.36(20)           \\
$\rho$, g cm$^{-3}$  & $\dots$    & $\dots$        & $\dots$      & 0.36(8)           \\
$g$, 10$^3$ cm s$^{-2}$ & $\dots$ & $\dots$        & $\dots$      & 0.99(46)         \\
%-------------------------------------------------------------------------------------
\hline
\end{tabular}
\tablefoot{
$^{[1]}$ \citet{but06};
$^{[2]}$ \citet{mca10};
$^{[3]}$ \citet{cur11};
$^{[4]}$ this paper;
$^{[5]}$ fixed parameter;
$^{[6]}$ JD 2,450,000+;
$^{[7]}$ $T_{\rm p}$ = 1800.805 and 4.749, $T_{\rm t}$ = 1803.021 and 6.965, to compare with $[1]$ and $[3]$, respectively;
$^{[8]}$ $\Omega$ can also be 56\degr\ due to the 180\degr\ ambiguity.
Errors corresponding to the last digits of the values are given in parentheses.
}
\end{table*}

We emphasize that for non-transiting planets  polarimetry provides a unique opportunity to directly evaluate the orbit inclination $i$ and $\Omega$. Moreover, it is possible to distinguish between inclinations smaller and larger than 90\degr, which is not possible from transit data. %Our definitions of the orbital parameters are given in FB10.
In our model (FB10) the inclination is defined in such a way that the planet revolves counterclockwise as projected on the sky for $0\degr\le i<90\degr$ and clockwise for $90\degr<i\le180\degr$. Further, $\Omega$ varies from 0\degr\ to 360\degr\ starting from the north and increases via east, south, and west. Note that there is ambiguity of 180\degr in the $\Omega$ value, which is due to the intrinsic property of Stokes $q$ and $u$ to remain unchanged under the rotation by 180\degr. The periastron longitude $\omega$ varies from 0\degr\ to 360\degr\ starting at the ascending node and following first the ascending part of the orbit.

\begin{figure}
\centering
\resizebox{\hsize}{!}{\includegraphics{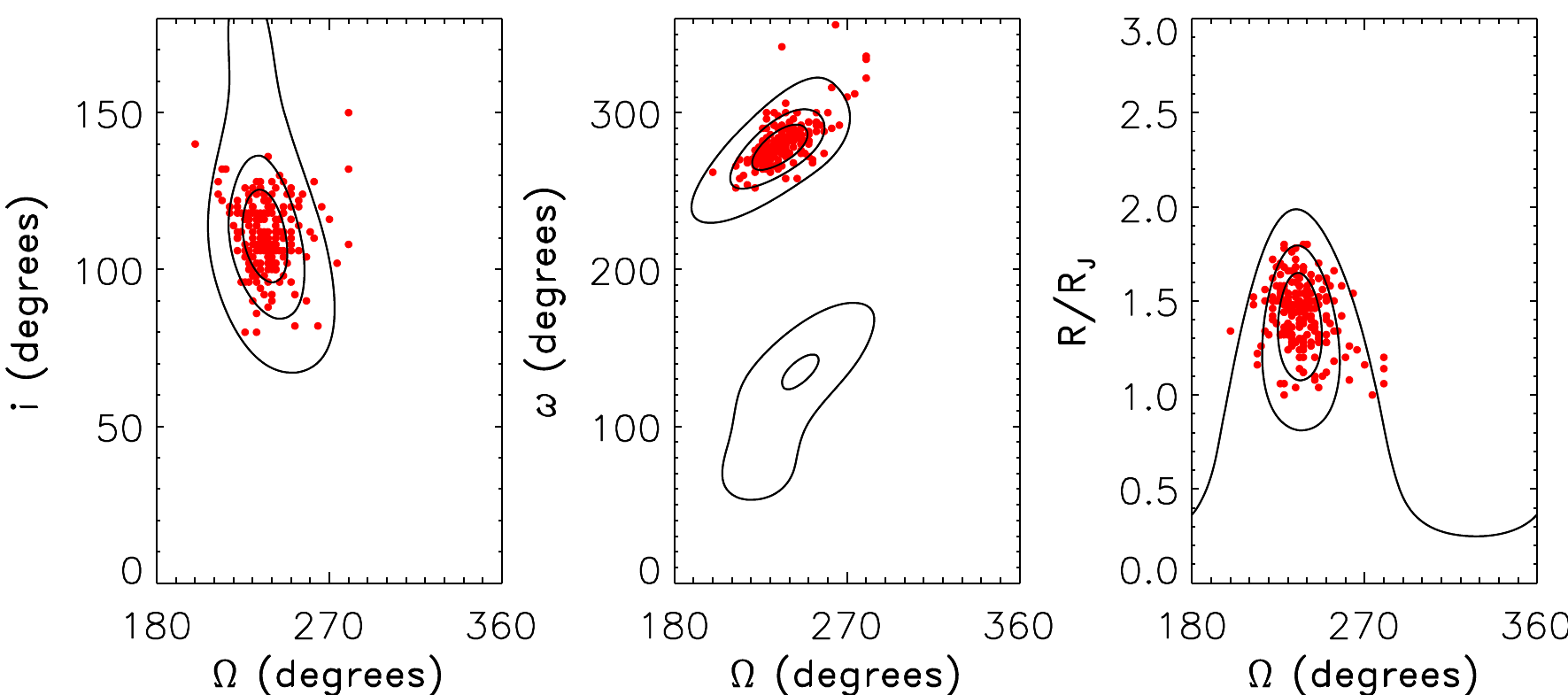}}
\caption{The $\chi^2$ contours of the best-fit solution for $\upsilon$~And~b in the $UBV$ bands simultaneously (solid lines). The three contours are shown for the confidence levels 68.3\%, 90.0\%, and 99.0\%. Due to the intrinsic ambiguity in $\Omega$ only one of the two equal minima separated by 180\degr\ is shown in each plot. The Monte Carlo sample solutions are shown with dots. }
\label{fig:chi}
\end{figure}

The four model parameters were first estimated from all $UBV$ measurements simultaneously, since this provided the highest accuracy of the orbital solution (31 degrees of freedom). For each spectroscopic solution we found practically the same values of $i$, $\Omega$, and $R_{\rm RL}$. The values of $\omega$ were found in agreement with the corresponding spectroscopic estimates. Their differences between the sets are due to differences in $P$, $T_{\rm t}$, and $e$. The solution with the smallest standard deviations was found with parameters by \citet{mca10} who combined spectroscopic and astrometric measurements for the planets b, c, and d. Our solution is presented in the fourth column of Table~\ref{tab:par}, and the fits are shown in the lower panels of Figure~\ref{fig:pol}. The errors of the parameters were estimated using Monte Carlo simulations as described by B08. The $\chi^2$ contours of this solution and results of the Monte Carlo simulations are shown in Figure~\ref{fig:chi}. 
The Monte Carlo tests were concentrated near the $\chi^2$ minimum for all parameters, which indicates that our fit to the data is robust to errors. 

\begin{figure}
\centering
\resizebox{7cm}{!}{\includegraphics{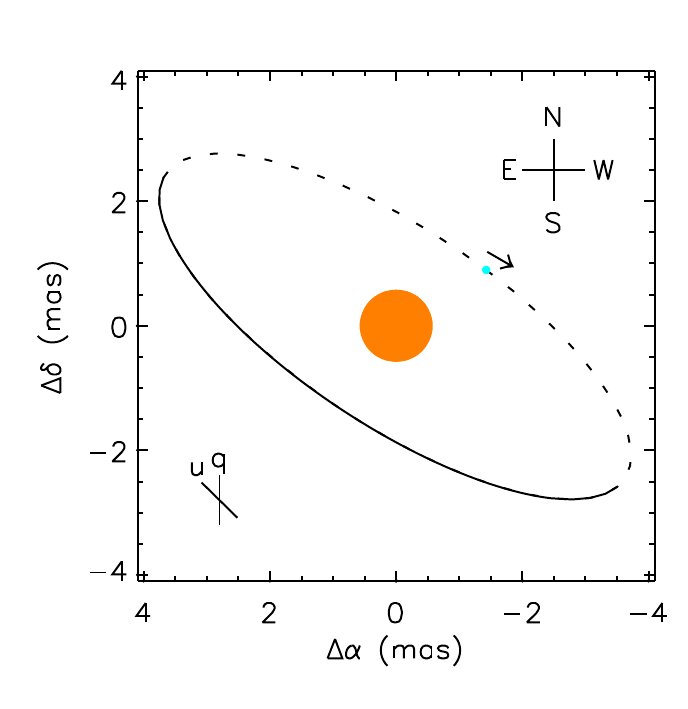}}
\caption{The orbit of $\upsilon$~And~b as projected on the sky. Solid and dashed lines indicate parts of the orbit in front of and behind the sky plane, respectively. The circle in the center depicts the star and the smaller circle on the orbit the planet. The direction of the orbital motion is indicated by the arrow. The reconstruction is made for the orbit parameters listed in the last column of Table~\ref{tab:par}, stellar radius $R_*/R_\odot=1.631$ \citep{bai08}, and distance $d$=13.47\,pc. The planet is positioned at the periastron. If $\Omega$=56\degr, the parts of the orbit in front of and behind the sky plane are to be swapped together with the periastron position but the direction of the orbital motion is preserved. Positive directions of Stokes $q$ and $u$ and orientation on the sky are also shown.
}
\label{fig:orb}
\end{figure}

The deduced epochs for planet b agree well with those by \citet{but06} and \citet{mca10}, but differ slightly from estimates by \citet{cur11} whose solution includes four planets and a trend but is based on spectroscopy only. The inclination of the planetary orbit found from polarimetry is close to an indirect estimate of the inclination of the stellar rotation axis deduced from the stellar rotation periods and projected rotation velocity: 58\degr$\pm$8\degr\ from the period of 7.3 days \citep{sim10} and 68\degr\ from the period 7.88 days (Sect.~\ref{sec:pol}). In fact, their complementary values 180\degr--58\degr=122\degr\ and 180\degr--68\degr=112\degr\ should be considered. Thus, the assumption that the stellar rotation axis is aligned with the normal to the planetary orbit made by \citet{sim10} seems quite good for $\upsilon$~And~b, but is not true for the planets c and d in this system according to astrometric measurements \citep{mca10}. The found inclination also satisfies the lower limit $i\ge28\degr$ deduced from infrared flux variations \citep{cross10} and is outside the limits $i=90\degr\pm8\degr$ implying transits. It is interesting that $\Omega$=236\degr of planet b is the same as for planet c \citep[$237\degr\pm7\degr$,][]{mca10} and that $\omega$ of planets b, c, and d \citep{cur11} are very close to each other too. This indicates that the apsidal resonance known for planets c and d includes also planet b, which provides further constraints on the dynamical history of this system.

The full set of the orbital parameters of $\upsilon$~And~b allows us to depict its orbit as projected on the sky plane and to indicate the direction of the orbital motion. This is shown in Figure~\ref{fig:orb}. Such a plot is useful for future direct studies of the system as the planet coordinates can be predicted quite accurately.

%***********************************************************************
\section{Physical properties of $\upsilon$~And~b}\label{sec:atm}
%***********************************************************************

Knowing the inclination of the orbit allows evaluation of a true mass of the planet $0.74 M_{\rm J}$.  This combined with the radius of $1.36 R_{\rm J}$ leads to quite low mean density and surface gravity (Table~\ref{tab:par}), which identify $\upsilon$~And~b as an inflated, low density planet \citep[e.g.,][]{bar10}. For instance, it is almost a twin to HAT-P-9b \citep{shporer09} and is very similar to HD209458b \citep{mazeh00}, WASP-1b \citep{cc07}, and HAT-P-1b \citep{bak07}, to name a few. Interestingly, they all orbit F7--G0 dwarfs at 0.05--0.06\,AU. Note also that about a quarter of known transiting planets have mean density lower than 0.4\,g\,cm$^{-3}$. Our 3$\sigma$ upper limit on the surface gravity $g<2.3\times10^3$\,cm\,s$^{-2}$ for $\upsilon$~And~b is consistent with the limit $g<2.1\times10^3$\,cm\,s$^{-2}$ deduced from 24\,$\mu$m flux variations \citep{cross10}.

These and earlier IR flux measurements by \citet{har06} revealed a large temperature contrast between day and night sides of the planet. Assuming $R(24\mu{\rm m})=1.3 R_{\rm J}$ and zero albedo, Crossfield et al.\ obtained the contrast $\Delta T\ge900$\,K and the lower limit on the orbit inclination of 28\degr. Using our inclination value of 111\degr, we can evaluate the contrast $\Delta T\sim1000$\,K. The assumption on zero albedo at 24\,$\mu$m agrees with our conclusion that it diminishes strongly toward the red (see below).

To investigate wavelength dependence of polarization and planet properties, we also modeled each $UBV$ passband separately while having fixed all other parameters but the radius. We obtained $R_{\rm RL}=1.34\pm0.35$, $1.50\pm0.28$, and $1.00\pm0.40$ $R_{\rm J}$ for $U$, $B$, and $V$, respectively. The corresponding fits are shown in the upper panels of Figure~\ref{fig:pol}. Even though we obtained a statistically significant signal only in $B$, the other bands provide useful limits at the corresponding wavelengths. An estimate of the polarimetric signal in red wavelengths ($RI$ bands) can be inferred from the upper limit by \citet{luc09} for $\tau$~Boo~b, when scaling it to the orbit size of $\upsilon$~And~b. This resulted in the polarization amplitude of $<3.3\times10^{-5}$ and $R_{\rm RL}(RI)<0.36 R_{\rm J}$. 

\begin{figure}
\centering
\resizebox{\hsize}{!}{\includegraphics{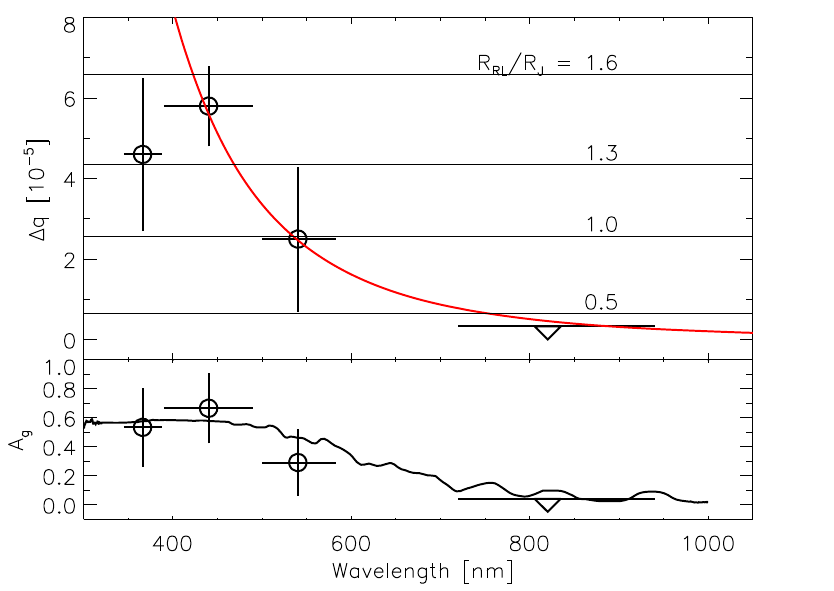}}
\caption{Polarimetric amplitudes $\Delta q$ (top) and geometrical albedo (bottom) measured in various passbands. Our new data are shown by open circles and the upper limit for $\tau$~Boo~b \citep{luc09} scaled to the orbit of $\upsilon$~And~b by a triangle. Horizontal bars show the passbands' widths. 
Upper panel:
thin horizontal lines indicate $\Delta q$ for different $R_{\rm RL}$/$R_{\rm J}$; 
solid (red) curve is the Rayleigh law $\Delta q = 2.1\times10^{11}\lambda^{-4}$ scaled to fit the $BVRI$ data.
Lower panel:
solid curve is the geometrical albedo of Neptune \citep{karkoschka1994} smoothed by a 100\,nm boxcar.
Our data indicate the dominance of Rayleigh scattering in the planetary atmosphere. 
}
\label{fig:amp}
\end{figure}

Similar to HD189733b (B11), the $BVRI$ polarization amplitudes indicate the dominance of Rayleigh scattering in the optical, as they can be approximated with a Rayleigh law (Fig.~\ref{fig:amp}, upper panel). The upper limit in $U$ is however significantly lower than this relation and even smaller than in $B$. As pointed out by B11, this can be a signature of additional opacity in the near UV which is perhaps stronger in $\upsilon$~And~b than in HD189733b. Taking into account that $\upsilon$~And~b is of lower density (more inflated) and orbits a hotter and metal-rich star ($[{\rm Fe/H}]\sim0.1$), this is in accord with the suggestion by \citet{bur07,bur08} that inflated planets may have larger optical opacities (at higher altitudes) resulting in temperature inversion (stratosphere) and higher infrared fluxes. One possible source of such opacity is absorption in TiO and VO bands \citep[e.g.,][]{fort08}. Our data indicate that there may be an additional source in the UV, e.g., absorption by a haze of photolysis products or Raman scattering. Being confined to the UV, this additional opacity may play a significant role in heating of the planet.

The large range of $R_{\rm RL}$ in different passbands indicates a strong wavelength dependence of the geometrical albedo, as discussed by B11. Assuming a maximum physical radius of the planet of 1.5\,$R_{\rm J}$, we can evaluate geometrical albedos $A_{\rm g}$ = 0.53$\pm$0.27, 0.67$\pm$0.24, 0.29$\pm$0.23, $<$0.04, in $U,B,V,RI$, respectively. These are plotted in Figure~\ref{fig:amp} (lower panel). Then, the average albedo for the range of 300--500\,nm is 0.60$\pm$0.18, and for the entire optical range is $\sim$0.3. Similar to HD189733b, we find a resemblance of $A_{\rm g}(\lambda)$ to that of Neptune, whose albedo is defined by scattering in the blue and by absorption in the red \citep{sromovsky2005}.  

A marginal spectral signature of the reflected light in the wavelength range 380--650\,nm was reported by \citet{cc02} at a projected orbital velocity amplitude of the planet 132\,km\,s$^{-1}$. Assuming orbital inclination of 70\degr--80\degr and flat geometrical albedo of 0.42 \citep[Class V models,][]{sud00}, they estimated the planet radius of 1.34$\pm$0.17\,$R_{\rm J}$. These assumed and deduced parameters appear very close to our results. For instance, our orbital parameters predict the velocity amplitude of the planet to be 138\,km\,s$^{-1}$. Thus, the Class V models with highly-reflecting clouds at lower pressure altitudes apparently hold for the $\upsilon$~And~b atmosphere (at least in the blue). This resembles our preliminary model atmosphere for HD189733b \citep{berd11spw6} with a high-altitude condensate layer which can reproduce the wavelength dependence of polarization and albedo if average particle size is $\sim$20\,nm. These clouds are highly reflective in the blue due to the small particle size. 

%***********************************************************************
\section{Geometrical albedos of hot Jupiters}\label{sec:alb}
%***********************************************************************

\begin{figure*}
\centering
\resizebox{\hsize}{!}{\includegraphics{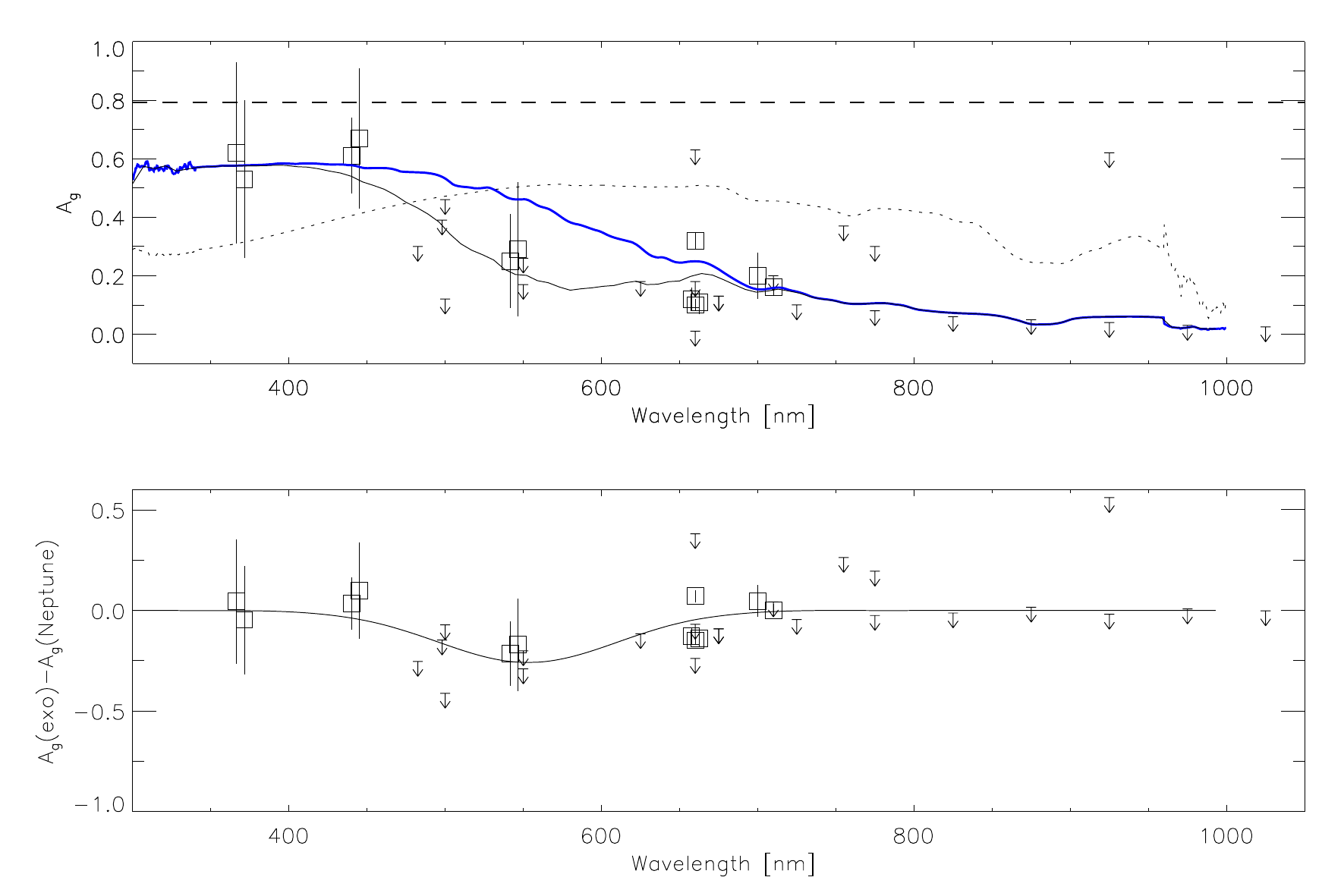}}
\caption{Geometrical albedo of hot Jupiters measured in various passbands (from Table~\ref{tab:alb}). Upper limits are shown by downward arrows. Some symbols are slightly moved from their central wavelength for clarity. In the upper panel, solid thick (blue) and dotted curves are the geometrical albedos of Neptune and Jupiter, respectively \citep{karkoschka1994}, smoothed by a 100\,nm boxcar. Dashed horizontal line indicates the albedo 0.791 of a semiinfinite conservative Rayleigh scattering atmosphere. Thin solid line is a modified Neptune albedo taking into account an additional absorption at 470--630 nm shown in the lower panel.
}
\label{fig:alb}
\end{figure*}

Detecting reflected light from exoplanets has been a challenging task since their discovery. First attempts were to extract a spectral signature from the observed spectrum of the star~+~planet system by comparing it to a modelled stellar spectrum, under the assumption that the reflected spectrum is a copy of the stellar spectrum \citep[e.g.,][]{cha99,lei03a}. These provided upper limits on the optical reflected flux for some hot Jupiters. For transiting systems, detecting reflected light can constrain the geometrical albedo $A_{\rm g}$ of the planet at a given wavelength. For instance, high-precision optical photometry in broad passbands with space telescopes MOST, CoRoT, and {\it Kepler} enabled detection or constrained secondary eclipses and light curves effects from planetary phases. Modeling these light curves, while neglecting or evaluating the contribution due to planet's thermal radiation, resulted in estimates of albedos of transiting hot Jupiters.

It is interesting to compare the polarimetric albedos of $\upsilon$~And~b and HD189733b with those of other hot Jupiters. We collected known estimates of $A_{\rm g}$ in Table~\ref{tab:alb} together with other releveant data on the planets and host stars, as provided by the exoplanet.eu database and references therein \citep{sch11}. The difference to our data is that passbands of the space-based measurements are significantly wider and centred in red wavelengths, and our albedo estimates limited to the range of 350--450 nm are unique. Also, measurements in several passbands for the same planet are still rare in the visible. 

\begin{table*}
\caption{Geometrical albedos of hot Jupiters.}
\label{tab:alb}
\centering
\begin{tabular}{lccccccccl}
\hline\hline
Planet & $A_{\rm g}$ & Passband & $R_{\rm p}$ & $M_{\rm p}$ & $\rho$ & $a$ & $T{_*}$ & $R_*$ & Method$^+$, \\
   &   & [nm] & [$R_{\rm J}$] & [$M_{\rm J}$] & [g cm$^{-3}$] & [AU] & [K] & [$R_\odot$]     & Reference\\
\hline
%-----------------------------------------------------------------------------------
\multicolumn{8}{c}{} \\
HD189733b &  0.62$\pm$0.31 & 345--388 & 1.178 &  1.138 & 0.86 & 0.03142 &  4980 & 0.788 & Pol, [1] \\
 "        &  0.61$\pm$0.13 & 390--490 &       &        &      &         &       &       &   "      \\ 
 "        &  0.28$\pm$0.16 & 500--583 &       &        &      &         &       &       &   "      \\
 "        & $<$0.26        & 450--650 &       &        &      &         &       &       & Pol, [3]  \\
$\upsilon$ And b & 0.53$\pm$0.27& 345--388 & 1.35  & 0.74 & 0.37 & 0.059 & 6212 & 1.631 & Pol, [2] \\
 "        &  0.67$\pm$0.24 & 390--490 &       &        &      &         &       &       &   "      \\ 
 "        &  0.29$\pm$0.23 & 500--490 &       &        &      &         &       &       &   "   \\
HD209458b & $<0.17$        & 400--700 & 1.38  &  0.714 & 0.34 & 0.04747 &  6075 & 1.146 & Ph, [4]  \\
CoRoT-1b  &  0.20$\pm$0.08 & 400--1000& 1.49  &  1.03  & 0.39 & 0.0254  &  5950 & 1.11  & Ph, [5] \\
 "        &  $<$0.2        & 560--860 &       &        &      &         &       &       & Ph, [6] \\
CoRoT-2b  &  0.16$\pm$0.03 & 560--860 & 1.465 &  3.31  & 1.31 & 0.0281  &  5625 & 0.902 & Ph, [7] \\
HATP-7b   &  $<$0.18       & 423--897 & 1.421 &  1.8   & 0.78 & 0.0379  &  6350 & 1.84  & Ph, [9] \\
 "        &  $<$0.13       & 350--1000&       &        &      &         &       &       & Ph, [8] \\
TrES-2b   &  $<$0.01       & 423--897 & 1.169 &  1.253 & 0.97 & 0.03556 &  5850 & 1.0   & Ph, [10] \\
TrES-3b		&  $<$1.07       & 550--700 & 1.305 &  1.91  & 1.07 & 0.0226  &  5720 & 0.813 & Ph, [11] \\
"      		&  $<$0.30       & 700--850 &       &        &      &         &       &       & "        \\
"      		&  $<$0.62       & 850--1000&       &        &      &         &       &       & "        \\
Kepler-5b &  0.12$\pm$0.04 & 423--897 & 1.431 &  2.114 & 0.90 & 0.05064 &  6297 & 1.793 & Ph, [12,13] \\
Kepler-6b	&  0.11$\pm$0.04 & 423--897 & 1.323 &  0.669 & 0.36 & 0.04567 &  5647 & 1.391 & Ph, [13] \\
Kepler-7b &  0.32$\pm$0.03 & 423--897 & 1.614 &  0.433 & 0.13 & 0.06246 &  5933 & 2.02  & Ph, [12,14]\\ 
Kepler-8b &  $<$0.63       & 423--897 & 1.419 & 0.60   & 0.26 & 0.0483  & 6213  & 1.486 & Ph, [12]  \\
Kepler-17b&  0.10$\pm$0.02 & 423--897 & 1.31  &  2.45  & 1.35 & 0.02591 & 5630  & 1.019 & Ph, [15] \\
$\tau$ Boo b &  $<$0.3     & 466--499 & (1.2)$^\times$ & $>$3.9  & 2.80 & 0.046 &  6309 & 1.331 & Sp, [16] \\
"            &  $<$0.39    & 385--611 & (1.2)$^\times$ &         &      &       &       &       & Sp, [17,20]\\
"            &  $<$0.37    & 590--920& (1.2)$^\times$  &         &      &       &       &       & Pol, [21]\\
HD75289A b&  $<$0.12       & 400--900 & (1.6)$^\times$ & $>$0.42 & 0.24 & 0.046 &  6120 & 1.25  & Sp, [18] \\
"         &  $<$0.46       & 400--900 & (1.2)$^\times$ &         &      &       &       &       & Sp, [19] \\
\hline
\end{tabular}
\tablefoot{
$^+$ Method to evaluate albedo: Pol -- polarimetry, Ph - photometry, Sp -- spectroscopy.
$^\times$ Assumed values.
References: 
[1] \citet{berd11}; 
[2] this paper;
[3] \citet{wik09};
[4] \citet{row08};
[5] \citet{alo09};
[6] \citet{sne09};
[7] \citet{sne10};
[8] \citet{chr10};
[9] \citet{wel10};
[10] \citet{kip11b};
[11] \citet{win08};
[12] \citet{kip11};
[13] \citet{des11a};
[14] \citet{dem11};
[15] \cite{des11b};
[16] \citet{cha99};
[17] \citet{lei03a};
[18] \citet{lei03b};
[19] \citet{rod08};
[20] \citet{rod10};
[21] \citet{luc09};
}
\end{table*}

This sample is rather homogeneous in terms of planetary and stellar parameters. In particular, all but one (HD189733b) planets are orbiting late F -- early G stars and at a distance smaller than 0.062 AU. Therefore, it makes sense to characterize this sample with average parameters:
$\langle T_*\rangle = 5930$~K, 
$\langle a \rangle = 0.04$~AU, 
$\langle R_{\rm p} \rangle = 1.4 R_{\rm J}$,
$\langle M_{\rm p} \rangle = 1.5 M_{\rm J}$, 
$\langle \rho \rangle = 0.8 $ g sm$^{-3}$, and
$\langle A_{\rm g} \rangle = 0.2$ (average over all optical wavelengths).
One can consider these averages as typical for a hot Jupiter in the vicinity of an early G dwarf. 

We find no significant correlations of the albedo with other planetary and stellar parameters. This can be in part due to large differences in the passbands. However, the wavelength dependence of the geometric albedo is of great interest, as it can well characterize the planetary atmosphere.

In Fig.~\ref{fig:alb}, we plot the available albedo estimates at the central wavelengths of the passpands. Upper limits are plotted as downward arrows. One can see a clear tendency for the albedo to decrease toward the red: from 0.6 in the near UV to essentially zero in the near IR. A few exceptionally high upper limits in the red are probably due to either very low signal-to-noise ratio or contamination by thermal emission. 

A comparison with the albedo of the Solar system giant planets is educative. Earlier we concluded that the polarimetric albedos of the two hot Jupiters seem to be similar to the albedo of Neptune. The measurements by CoRoT and {\rm Kepler} do not contradict this observation. Moreover, it is rather obvious that practically all measurements strongly differ from those of the Jupiter's albedo, which is dimmer in the near UV and brighter in the red. Hence, Neptune's albedo is a more suitable reference for comparison with hot Jupiters. 

As pointed out above, the albedo of the cold Neptune is mainly shaped by Rayleigh and Raman scattering on a high-altitude haze and H$_2$ in the blue, and methan absorption in the red \citep{sromovsky2005}. In hot Jupiter atmospheres, H$_2$ is still main scatterer, together with \ion{H}{i}, while absorption in the red is mainly due to TiO and H$_2$O \citep[e.g.,][]{sud00,bur08,fort08,dem11}. These are perhaps the main opacities influencing the overall shape of the hot Jupiter albedo. Model predicts that the presence of high-altitude haze can significantly increase the albedo. This is revealed in HD189733b with polarimetry \citep{berd08,berd11}, and transmission spectroscopy recently extended to the near UV  confirms the overall presence of haze and Rayleigh scattering \citep{pont08,lec08a,sing11}. This is in line with our preliminary semi-empirical model of HD189733b \citep{berd11spw6} which explains the observed polarization by the presence of 20 nm dust condensates in its atmosphere.

A closer inspection of the sample indicates that there is perhaps a systematic difference between exoplanets and Neptune, as shown in the lower panel of Fig.~\ref{fig:alb}. It extends over about 500--650 nm and reduces the albedo by $\sim$0.2. This dimming can be caused by absorption in the sodium resonance doublet (590 nm) and the TiO electronic band systems (450--650 nm). These species play no role in the Neptune atmosphere but are essential for hot Jupiters. A varying strength of the absorption can strongly influence broad-band optical albedo values. For instance, the very low optical albedo of HD209458b \citep{row08} can be due to an extreme strength of this absorption feature, which was also featured in the transmission spectroscopy by \citet{sing08}. They also detected an opacity increase towards shorter wavelenths which was interpreted as a signature of Rayleigh scattering \citep{lec08b}. Hence, also for HD209458b, we can expect somewhat higher albedo in the blue. In other planets, such as Kepler-7b, a lack of sodium absorption and/or presence of haze could promote high albedo \citep{dem11}.

%***********************************************************************
\section{Conclusions}\label{sec:dis}
%***********************************************************************

Our polarimetric study of $\upsilon$~And has unveiled many unique and important implications for the dynamics and evolution of its planetary system and for physical properties of the planet b and hot Jupiters in general.
\begin{itemize}
\item
Determining the inclination of the $\upsilon$~And~b orbit and evaluating its radius, for the first time, allowed us to correct the planet mass, constrain its density, and reveal its inflated nature. 
\item
The found periastron longitude indicates that the apsidal resonance known for planets c and d includes also planet b, which provides further constraints on their dynamical history.
\item 
The limits obtained on the geometrical albedo idicate reflective properties like in HD189733b. Both these hot Jupiters reflect most efficiently blue light and shine similar to Neptune in the Solar system. Their average optical albedo is close to that of Kepler-7b. It is possible that their high reflectivity is due to high-altitude haze. 
\item
Combining the polarimetric albedos with those evaluated from optical spectroscopy and photometry for a sample of planets, we obtained a unified picture of the geometrical albedo of an average hot Jupiter. Its albedo can be as high as 0.6 in the near UV and close to zero at wavelengths longer than 700 nm. A broad absorption centred at 550 nm, probably due to TiO and \ion{Na}{i}, can strongly influence an average optical albedo of hot Jupiters. 
\end{itemize}

Our results define clear prospects for follow up studies of hot Jupiters. First, determinig albedo in the blue ($\lambda<500$ nm) for a larger sample of planets is crucial for clarifying whether the high reflectivity in the blue is a typical feature of hot Jupiters. Second, new self-consistent model atmospheres of hot Jupiters should be developed to explain such high reflectivity and to constrain composition of their atmospheres, especially at high altitudes. Finally, polarimetric constraints on planet reflectivity are essential for separating reflected light from thermal emission in the flux measured through secondary eclipse photometry.

%---------------------------------------------------------------------------
\begin{acknowledgements}
S.V.B. acknowledges the EURYI (European Young Investigator) Award provided 
by the European Science Foundation (see www.esf.org/euryi) and SNF
grant PE002-104552. This work is also supported by the Academy of Finland, grant 115417. 
Observations were made with the Nordic Optical Telescope, La Palma, Spain, operated on the island of La Palma jointly by Denmark, Finland, Iceland, Norway, and Sweden, in the Spanish Observatorio del Roque de los Muchachos of the Instituto de Astrofisica de Canarias.
\end{acknowledgements}

\end{document}